*Article*

# Still Unsolved High-Performance Computing Challenges for up to Pre-Petascale Homogeneous Supercomputers

**Mindaugas Macernis[1]*, Vaidotas Mickus[2], Janne Ahonen[3], Laurynas Diska[1], Jonas Franukevicius[1] and Juozas Sulskus[2]**

[1] Institute of Chemical Physics, Faculty of Physics, Vilnius University, Sauletekio 9-III, 10222 Vilnius, Lithuania;
[2] Duvait UAB, Konstitucijos pr. 21A, LT-08130 Vilnius, Lithuania;
[3] Bull Baltija UAB (Part of ATOS group), Konstitucijos pr. 21A, LT-08130 Vilnius, Lithuania;
* Correspondence: mindaugas.macernis@ff.vu.lt; Tel.: +370 5 223 4659

**Abstract:** Pre-exascale High Performance Computers (HPC) can reach more than 400 Pflop/s real performance according the HPLinpack benchmarks. For nanoscience and quantum biology there are requirements for those program codes based on quantum physics algorithms which is difficult to ideally parallelize. Such parallel codes reach their limitations at terascale performance clusters. The standard Amdahl's law suggestions for code parallelization complicates focusing and planning for the next step the parallel code developments. In this report we focused on a three key applications domain which are highly parallelizable: HPC benchmarks, quantum computing simulators and Car−Parinello molecular dynamics. According the results we summarize the Amdahl's Law & Parallel Speedup performance achievements with supercomputer with pre-petascale homogeneous HPC hardware. We conclude as an universal computer the pre-petascale supercomputing performance homogeneous hardware still has the basic challenges which must be addressed by the researchers or developer in order efficiently to use them.

**Keywords:** high performance computing; quantum chemistry; molecular dynamics; speed-up; quantum computer simulators

## 1. Introduction

Today two new emerging fields are appearing quantum biology [1-3] and quantum computing [4-6]. The quantum effects in complex biological systems problems are based on quantum mechanics and the computation are at scale of the large compactional resources what means there are a need for the massive parallelization which is a challenge itself [7]. The Amdahl's Law allows to predict the limitations for classical code with classical computer hardware [8-11] but the quantum computer is expected to give new breakthrough in modeling exponential type tasks of quantum mechanics and biology[12, 13].

The multicore processor is mainstream processor architecture today and Amhdal's law revision is at interest in order better understand the upper limits of parallel code [8]. Also, the Amhdal's law is good for prediction of hardware computation limit or speedups at various scenarios such as communication costs[14], memory bandwidth[15], resource contention[10], impact of different workload types[11], power constraints[16] or other hardware scaling effects[9].

Linpack package primary was developed to collect the information on execution times to solve a system of linear equations [17]. This was very useful for researcher that he could estimate the required time to solve their matrix problem. The tool become more sophisticated, updated with parallel processing which resulted a collection of Fortran subroutines for solving various systems of linear equations[18]. Today the High-Performance Linpack (HPLinpack) benchmark become standard tool to measure the supercomputer performance and according this information the fastest 500 supercomputers are listed since 1993 in www.top500.org webpage [19].

Quantum computers can run some algorithms which scale better than their classical counterparts with difference as exponential scale-up [20]. In order to describe the state of $n$-bit classical register requires $n$ bits while an $n$ quantum bits (qubits) register requires $2^n$ complex numbers. The same remains for simulating a quantum computer with the classical machine. According it there is requirement exponentially costly resources with the respect to the number of qubits. Despite this fact, the studies for new quantum computing algorithms and architectures require the quantum computer simulations [4]. The universal quantum computing simulators in order to simulate all information of the full quantum state require the large computational resources in order to get results, especially the computer must have large Random access memory (RAM) size [4].

Computational studies are unique for studying various physical observables of the system while it allows to get additional information which is difficult to obtain from experiment. Computational modeling is important in many research areas such as chemical, physical and biological sciences. While various software and algorithms are very different there many problems which do not fit to standard computational speed-up predictions. Even the highly quality computational codes for parallel computer architectures has its specific limitations [7].

The focus on pre-petascale homogeneous hardware is no longer relevant - large scale computing systems are increasingly heterogeneous, incorporating components such as GPU accelerators [21]. Moreover, the heterogeneous systems have speed-up problems in order to use many GPUs which lacks the analysis in computational studies and it is provided in various internal tests and documentations (e.g. see https://ambermd.org/GPUHowTo.php). This leads the application algorithms relies more and more on accelerators leaving the "hard problem" calculations for the future quantum computer technologies [5, 6]. Amdahl's Law does not serve well as a performance prediction tool, as it is very coarse-grained and does not account for many aspects of system performance (inter-node communication is only one example).

In this paper we used HPLinpack to test and measure the performance of the homogenous architecture supercomputer with pre-petascale supercomputing performance homogeneous hardware. Here we examine the maximum possible simulations of fully described quantum computer qubits, we optimized and benchmarked the Car–Parinello molecular dynamics highly paralized codes which can be used to study for other quantum chemistry problems also. According collected data we discussed and described basic of Amhdal's law for code parallelizing.

## 2. Materials and Methods

### *2.1. Standard HPC benchmarks*

The High-Performance Linpack (HPLinpack) is software that solves random dense linear system in double precision arithmetic on distributed memory computers. The tool is useful for supercomputer testing and timing which is realized by quantitation the accuracy of the obtained solution. The major parameter of the HPLinpack is performance value which lets the researcher understand the limits of possible scalability that the parallel computation can be achieved. The HPLinpack requires Message Passing Interface (MPI) and Basic Linear Algebra Subprograms (BLAS).

In this study we used HPLinpack 2.3, version December2, 2018. We optimized the software by using BLAS library from the Intel 2020 version compilers. We used the standard communication over IB network for Message Passing Interface (MPI) which engineers of Bull/ATOS implemented into the supercomputer SLURM batch system.

We performed HPLinpack tests by repeating them several times in order to collects the maximums possible performances achievement. Every computing node we checked with the separate HPLinpack test and then step-by-step we increased the computing nodes for the whole system test. This procedure allows to check the maximum possible performance of the hardware and tune it.

*2.2. Quantum Molecular Dynamics Benchmarks*

Computational modeling, especially of quantum chemistry, realization is important in answering key scientific questions. According the Amdahl's Law and its Parallel Speedup problems the high-quality computational code requires to take advantage of massively parallel computer architectures. Such code is the NorthWest computational Chemistry (NWChem) software package developed in the W.R. Wiley Environmental Molecular Sciences Laboratory (EMSL) at the Pacific Northwest National Laboratory (PNNL) [7].

The same optimizations as for HPLinpack we used for NWChem 7.0.2 programing codes. Additionally, we noticed that by default NWchem compilations scripts lack optimization for Global Array (GA) library thus we updated it by adding the Intel(R) Xeon(R) Gold 6130 instructions.

Previously we performed and tuned the input files for the Car−Parinello Molecular Dynamics (CPMD) supermolecular approach calculations with up to 1.5 ps dynamics [22]. At first were tuned NWChem program code with Intel 2020 compilers and then we used CPMD codes for the NwChem benchmarking. The latter procedure was such: the first step was to change computing nodes by increasing their number at 16 ratios, and the second step was to perform computations with various different number computing nodes between 16 ratios.

*2.3. Banchmarks for large RAM*

The QuEST runs efficiently on all architectures available to a researcher what facilitate the seamless deployment of the researchers' code. This allows easily compare the performance of the different architectures and pick that which is suitable for their needed simulations.

The same optimizations as for HPLinpack we used for QuEST[4] programing codes. The QuEST code gathered information about SIMD parameters by default and we left them as it was due to fact that number of simulated qubits depends on Random access memory (RAM) size.

The most important Quantum computing algorithm is quantum furrier transfer (QFT) [4]. The QFT is used for various quantum algorithms as well as for quantum cryptography. Thus, we studied QuEST's algorithm calculations by trying to reach maximums possible qubit system for QFT algorithms.

*2.4. Optimizations*

There is various type of code optimizations which are implemented into hardware or software level. The most common hardware optimization accessible for researcher are processor and network architectures. Understanding the limitations and requirements for developed code is crucial in order to speed-up calculations. Moreover, the wrong executions can result so slow-down of the code.

Single-multiple-data (SIMD) execution can speed up at the processor (CPU) level. The common Intel's SIMD instructions are Advanced Vector Extensions (AVX), AVX2 or AVX-512 which can be used at compiler level with usage of careful code benchmarking. Additional parallelism can be achieved by taking advantage of multiple cores found in Central processing unit (CPU). This requires the code adaption for parallelization. The modern CPU supports Hyper-Threading (HT) technology for every physical core but the data transfer studies in the code must be done; otherwise the results can be the slow-down instead the speed-up. Next speed-up increase can be achieved by using Non-uniform memory access (NUMA) memory space, exchange of data in CPU's las level cache (LLC) with main memory.

While the single node parallelism is limited to CPU or GPU hardware the distributed architectures should consider the network architecture for communication between multiple computing nodes. The distributed architectures join thousands of cores and join large amount of the memory. So, there is a need to program all communication at the code level. The common standard is MPI for scientific type codes. The High-Performance Linpack (HPLinpack) tests allow better understand the hardware limitations and the research's possibilities in order to optimize the code for speed-up.

### *2.4. Hardware*

We evaluate the performance using "VU HPC" Saulėtekis location computing facilities, specifically supercomputer with the Sequana X1000 architecture in Vilnius University. For Message Passing Interface (MPI) which engineers of Bull/ATOS implemented into the supercomputer SLURM batch system.

The used computing nodes were from 1 to 143 where two of them are with 2x GPU Nvidia Volta 100 accelerators. The computing nodes had 32 physical Cores which can be used as 64 HT Cores. Each processors architecture is Intel(R) Xeon(R) Gold 6130 CPU @ 2.10GHz. Every computing node has 376GB RAM. Such architecture provides 12GB of RAM to 1 physical Core, or 6 GB of RAM for to 1 Logical Core (when using HT). The two computing nodes join 8 Nvidia Tesla V100-SX accelerators. The InfiniBand (IB) network is dedicated for communications in computations which has the rate 100 Gb/sec with 4X EDR. IB network is Bull/ATOS BXI technology with latencies less than 1 μs using MPI communication (see https://www.slideshare.net/insideHPC/bxi-bull-exascale-interconnect, slide 7).

## 3. Results

### *3.1. High-Performance Linpack Benchmarks*

The HPLinpack performance test is major tool in order inspect supercomputer and understand the limitations. At first together with the HPLinpack test is good understand the Amdahl's Law & Parallel Speedup limitations. We benchmarked every computing node separately with HPLinpack test (Figure 1). The same figure would be if the run would be performed for hundred times on one computing node also. As it can be seen the value has up to 10% changes at every test. However, in average the one computing node achieved the performance around Rmax 84% which is 1.8 TFlops. These results suggest the whole supercomputer theoretically cannot achieve larger value than Rmax 84%.

We achieved maximum value R=0.237 Pflops with 0.31 PFlops homogeneous HPC hardware by excluding GPU cores. The results varied at different running: the largest variation is at the performance peak starting from 128 compute nodes which can be seen in Figure 2. According data the performance degradation can be described as efficiency which is about 80% using up to 128 computing nodes. In similar cluster environments (DDR4 RAM, EDR InfiniBand network) we saw performance degradation down to 63% when running HPL Linpack on more than 128 compute nodes (Atos laboratory internal data, the efficiency was gathered with 240, 1000 and 1116 compute nodes). This is expected behavior because DDR4 RAM latency is up to 0.5 μs and InfiniBand latency is up to 1 μs. Summing up, most effective one calculation node count in environment is up to 128 nodes for effective global memory sharing between nodes using finely tuned MPI environment.

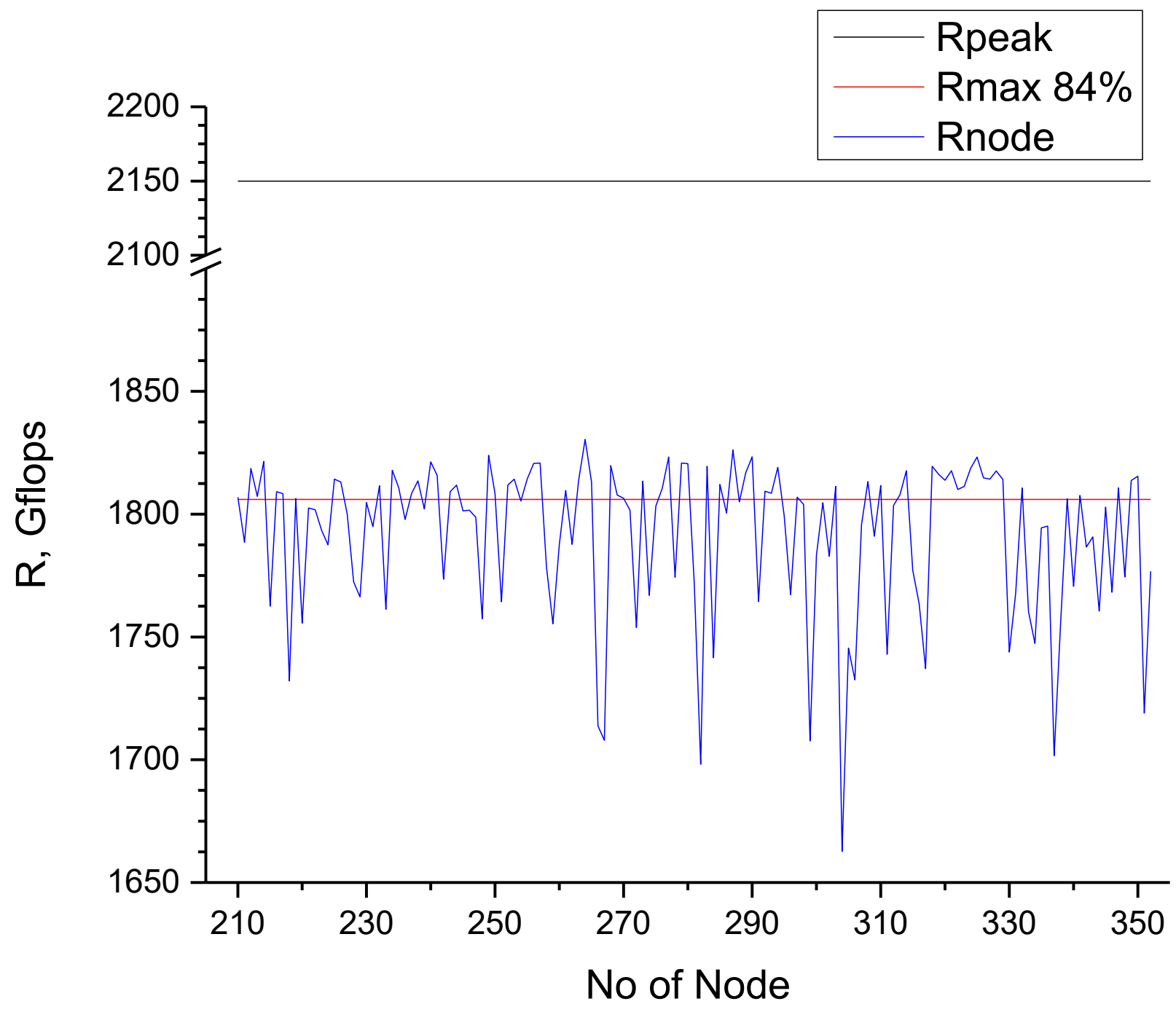

**Figure 1.** HPLinpack results: fix image of results for every computing node.

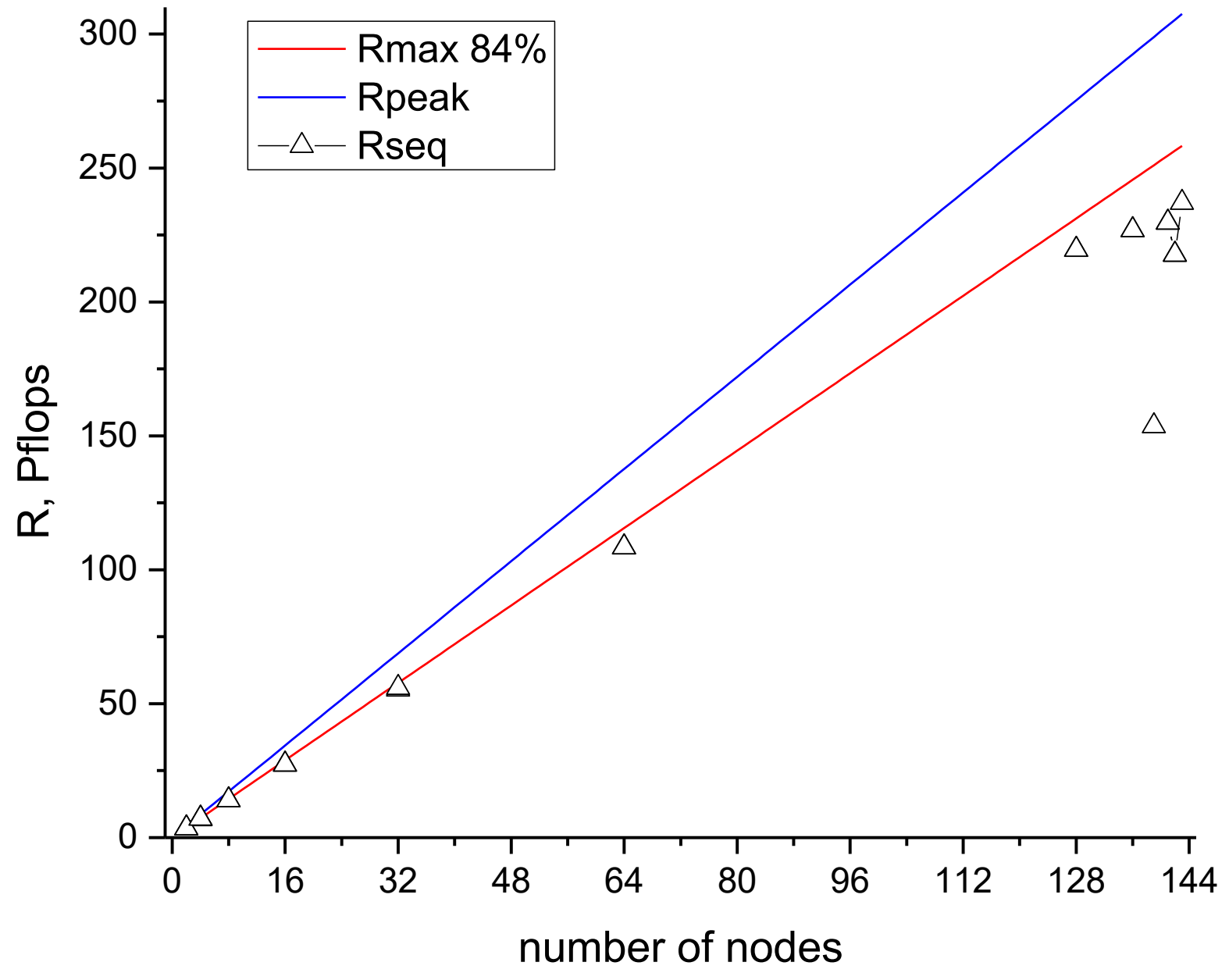

**Figure 2.** HPLinpack results: reached computing performance by increasing node numbers.

### *3.2. Qubit Number Benchmarks for Quantum Computers Simulations*

We successfully simulated fully described up to 40 qubits with the 128 computing nodes (Figure 3). We needed 64 computing nodes for the 39 qubit simulations. According this information it is clear that in order to simulate 41 qubits we would need twice number of the presented nodes. The scaling was ideal according the Random-access memory (RAM) consumption perspective: it required double amount of RAM for every addition simulated qubit. The computation time increases significantly when we simulated more than 30 qubits. Also, we should note the QuEST parallelization was related to minimal chosen computing node number, e.g the minimum 6 qubits must be simulated while using 2 computing nodes.

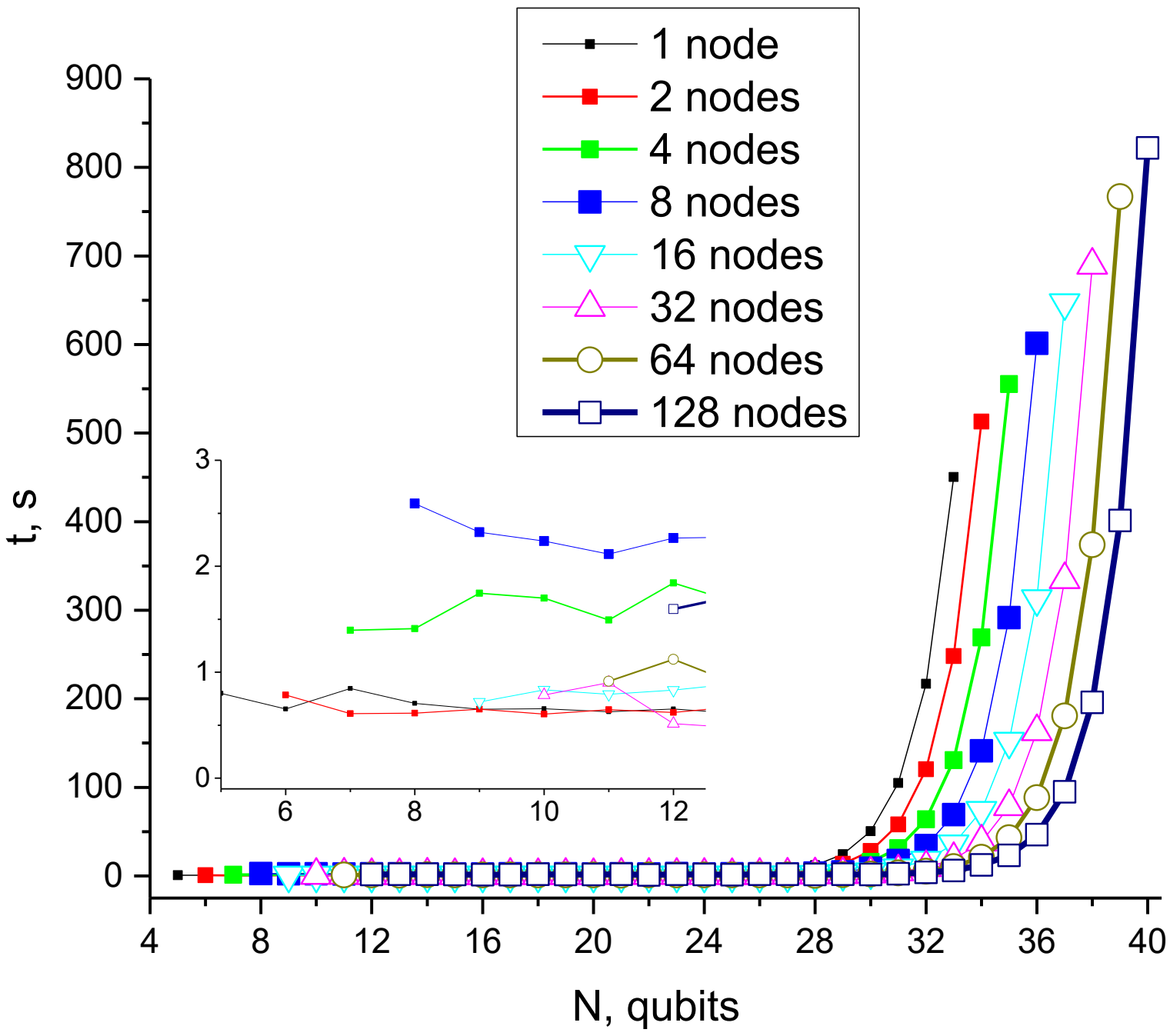

**Figure 3.** Simulations for maximum qubit number by changing the computing node number. The inset show the minimum number of qubits for small amount of computing nodes.

### *3.3. Car–Parinello Molecular Dynamics benchmarks*

In order to test real life job we chose job from previous run as it is described in ref. [22] . Previously the Car–Parinello molecular dynamics (CPMD) supermolecular approach calculations for 1.5 ps dynamics were calculating over one month with 8 computing nodes having AVX processor instructions over 40 GB/s IB network while the larger number of computing nodes drastically slowed down the computations. In order to measure the speed-up, the same job was restricted to the equilibrium tasks which takes 1000 MD steps. This allowed to achieve results over reasonable time for one computing node: 89.3 hours. Thus, we could expect the 1.5 ps simulation should take make about 3 months (what was true in ref [22]).

In Figure 4 we summarized CPMD speed-up timings with various number computing nodes which we used in benchmarks. Using 24 computing nodes the computation time decreased to 4.8 hours. The maximum speed-up we got with 96 computing nodes which was 2.53 hours. The 128 computing nodes shoed slightly slow down comparing to 96 computing nodes. The 1.5 ps simulation results could be achieved over 1 or 2 days. According ref [22] and our benchmarks we concluded there are huge correlations with the network I/O itself independently to the parallelization code itself.

Additionally, by analyzing speed-up we noticed the ratio of 16 importance for the program code scaling also. In all cases if the computations took more than 100 hours they were stopped. It is clearly seen that starting from 7 computing nodes the ratio of 8 becomes very important: the calculations took unreasonable long for running with 7, 15, 112 or 140 nodes. The same effect we got when the resources were allocated to every HT Core instead of the physical Core. This problem can not be attributed to the hardware issues but it should be linked to the code compilations itself.

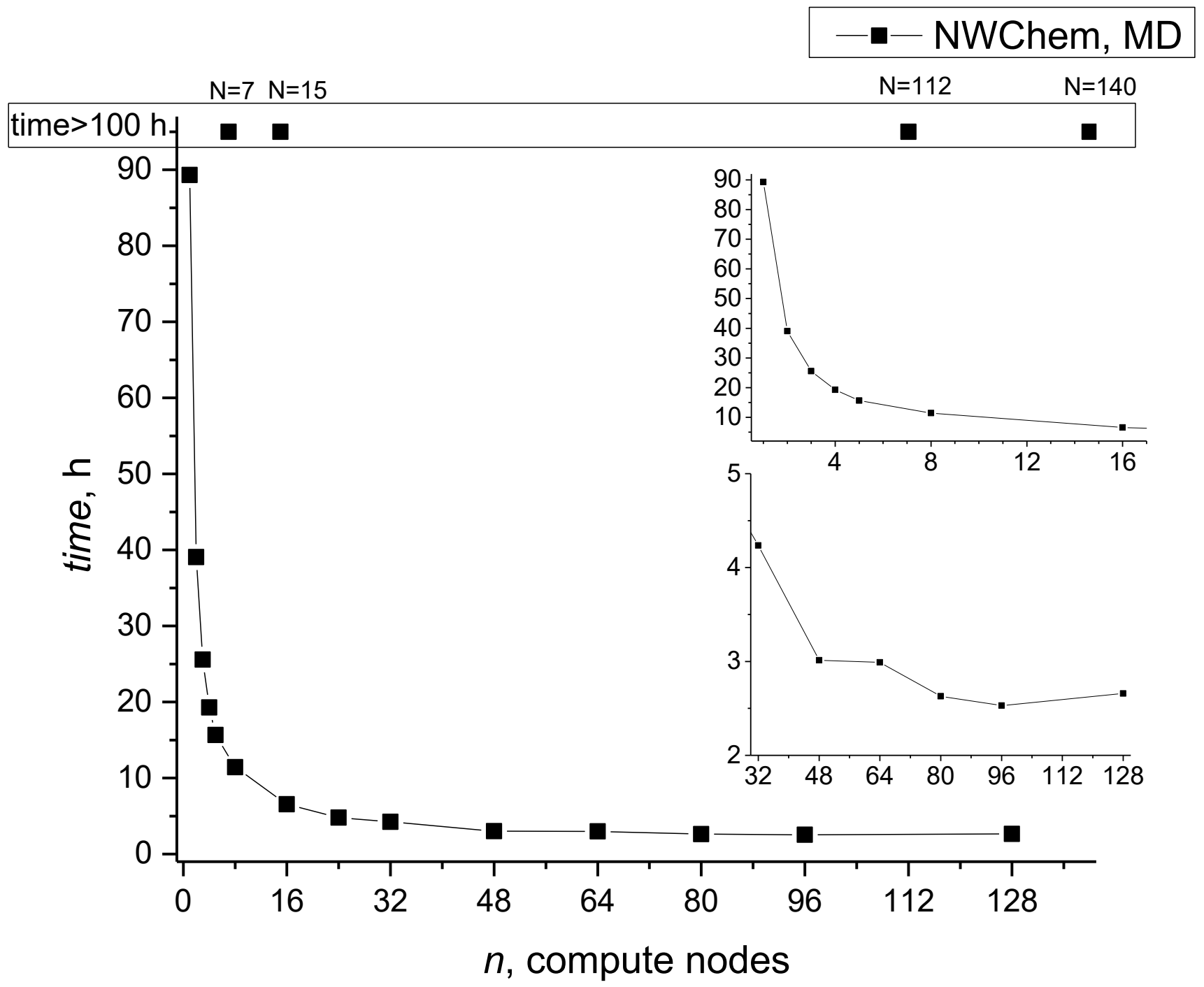

**Figure 4.** Car-Parrinello microcluster calculation equilibrium computational part by changing the computing node number. Insets: the zoomed data is shown in insets: top – the data at small amount of compute nodes; bottom – the data at largest amount of compute nodes.

## 4. Discussion

According the HPLinpack benchmarks it is clear that the researchers must consider the fact that even the highly parallelized code cannot achieve speed up due to many hardware initialization reasons (Figure 2). Even the same computing node can vary in timings which can be noticeable in very long computations, e.g. if we expect 5% of slow-down for 10 days calculations it will result to increase by additional half day calculations. We would recommend that the researcher should predict the computation time and he must continuously monitor if calculations run according his predictions, otherwise simply he should restart the job.

The quantum computer simulations took over 15 minutes for 40 qubits simulations while it run for several seconds with the lower number qubits. Thus, algorithms slow-down by thousands of times for the largest possible system. This means the small tested algorithms should run up to 1 minute with one or two computing nodes in order it could be tested with >30 qubits. Otherwise the calculations should take for unpredictable very long timings.

Highly paralyzed and optimized for massive parallelization CPMD program codes [7] may have simple computational problems as we notice the ration of 16 or 8 is very important for the speed-ups. Thus, the researcher should be careful by using large number of computing nodes. Moreover, the Amdahl's Law do not have prediction for such scaling problems. It is impossible to fit it according the calculation timings due to fact that they take too long and the problem is primary inside the programing code scaling algorithms. On other hand the network issues can be fitted to Amdahl's Law predictions.

### *4.1. Scale-Out Systems Affect Amdahl's Law*

Amdahl's Law & Parallel Speedup – do not include grid size effect which is clearly seen in GPU programing, Overloading or HT. Depending on data additional limitations can increase or decrease performance; scaling by ratio of 2 or 16 which is in many scenarios is crucial to get maximum performance if the code is quite well paralyzed.

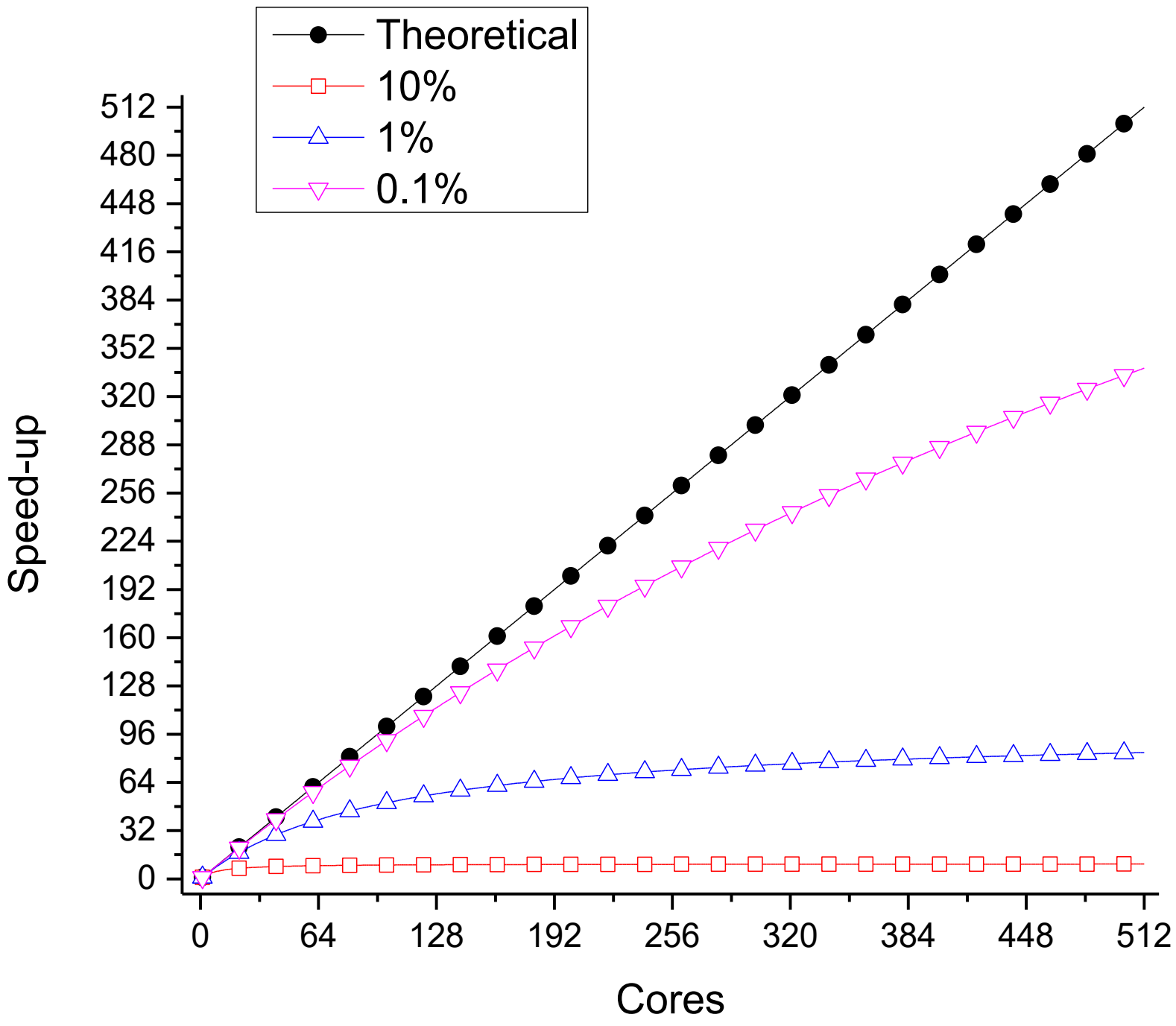

Fig. 5. Amdahl's Law: even 1% of serial code will result easily watchable speed-up upper limit at 64 or 128 cores/threads.

The basic Amdahl's law represented by formula compares the execution time of the serial and parallel parts of the program respectively with the $N$ number of processors. The Amdahl's law expression can be rewritten in simple form as percentage relation which only includes serial and parallel parts. Traditional Amdahl's law shows the upper limits with in the ideal scenarios (Fig. 5). This allow to predict how much of the code must be parallelized in order to have speedup, e. g. 1% serial code will result upper bound limit at about 128 cores or threads.

Traditional Amdahl's law does not involve various additional problems such as scaling according memory (RAM) amount, communication bandwidth and I/O latency limits between the computing nodes or competing processors Cores for the same hardware resources. For example, in ref. [23, 24] is described OpenACC computation limits by extending the Amdahl's law with additional parameters: the total execution time of a program was divided into the computation time and the memory access time. The computations threads can be assumed as dependent or independent which allow to describe speed-up and "speed-up bound" for resource-bound workload resulting the slowdown in large thread scenarios [8]. However, the traditional simple equations extension for the description of the hardware communication problems are related by extending some constant $k$ in its denominator.

By analyzing various parallelized software packages (not presented in the paper) we figure out that the same effect can be related to code communications realization itself. This means the constant $k$ value can be useful not only for hardware developments but also for the software code upper bond limits understanding. It is clear that CPU cores have at least two types of communications: inside the computing node or over the network interface. Thus, we suggested to message codes parallelization limits by describing the percentage from the ideal network communication:

$$\text{speedup} = \frac{1}{S + \frac{(1-S)}{N} + C\frac{N}{N_c}} \quad (1)$$

where S is percentage of serial code part of all the code, C is percentage for network communication idle and $N_c$ is core number in one node. This relation leads to relation that there is code slowdown with the large core number (Figure 6). The scenario with 512 Cores or other theoretical processes shows that in some cases communication issues must be improved by 100x times. Such description allows to have strategies how to optimize code by comparing parallel code and communication idle problems. As it can see in (Figure 6) in some cases the serial code conversion into parallel will not lead to speed-up with large number of cores: the code must be improved at communication parts inside parallel code between cores or threads level.

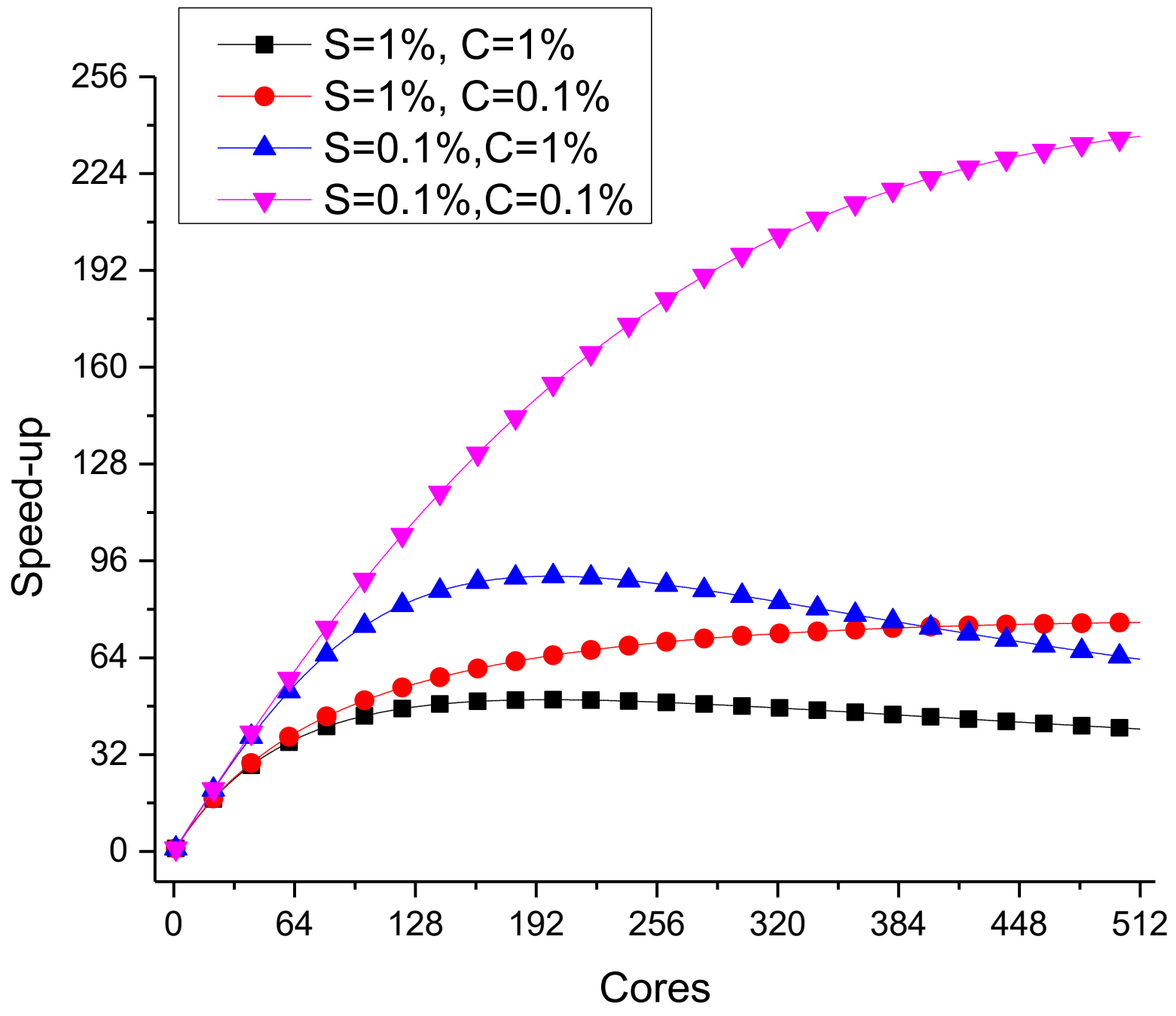

**Figure 6.** Amdahl's Law & Parallel Speedup: strategies to optimize code between parallel code and communication idle by comparing different scenario of optimizing serial code and network communication code parts. The blue line represents typical scaling problem for purely optimized parallel code.

## 5. Conclusions

In this paper, we described and analyzed pre-petascale supercomputing performance homogeneous hardware. Based on Intel Skylake type many-core processor the supercomputer has adopted other technologies, including integration thousands of cores, water cooling system, it provided peak performance over 0.237 PFlops of the HPLinpack tests. A number of key scientific applications have been optimized and benchmarked. The quantum computer simulators successfully simulated 40 qubits fully described system but the reasonable systems for the studies are for 30 qubits. CPMD computations provided that the real-life applications must be run carefully and the scaling ratios can be the key factor also. We updated the Amdahl's law formula with the communication description predictions inside parallel code which should help better to focus for the need either update the software algorithms or hardware network commutations.

**Author Contributions:** Conceptualization, M.M. and J.S.; methodology, M.M., V. M., J. F. and J. A.; software, M.M., V. M., J.F. and J. A.; validation, M.M., V. M. and J. A.; formal analysis, M. M., J.F. and L. D..; investigation, M.M.; resources, J.S. and M.M.; data curation, M.M. and V.M.; writing—original draft preparation, M.M.; writing—review and editing, J.S., L.D., V. M. and J. A.; visualization, M.M.; supervision, J.S.; project administration, M.M.; funding acquisition, M.M. All authors have read and agreed to the published version of the manuscript.

**Funding:** This research was funded by Research Council of Lithuania, grant number P-MIP-20-47.

**Institutional Review Board Statement:** Not applicable.

**Informed Consent Statement:** Not applicable.
**Data Availability Statement:** Not applicable.

**Acknowledgments:** Computations were performed on the supercomputer "VU HPC" Saulėtekis at Faculty of Physics in Vilnius University.

**Conflicts of Interest:** The authors declare no conflict of interest.